# Direct Reconstruction of Terahertz-driven Subcycle Electron Emission Dynamics

Jiakang Mao,[1,2,*] Yushan Zeng,[1,2,*] Hongyang Li,[1,3] Liwei Song,[1,2,†] Ye Tian,[1,2,‡] and Ruxin Li[1,2,4,§]

[1]State Key Laboratory of Ultra-intense laser Science and Technology, Shanghai Institute of Optics and Fine Mechanics (SIOM), Chinese Academy of Sciences (CAS), Shanghai 201800, China.

[2]Center of Materials Science and Optoelectronics Engineering, University of Chinese Academy of Sciences, Beijing 100049, China.

[3]School of Physics Science and Engineering, Tongji University, Shanghai 201210, China

[4]Zhangjiang Laboratory, Shanghai 201210, China

While field-driven electron emission is theoretically understood down to the subcycle regime, its direct experimental temporal characterization using long-wavelength terahertz (THz) fields remains elusive. Here, by driving a graphite tip with phase-stable quasi-single-cycle THz pulses, we reveal distinct subcycle electron emission dynamics including: (1) At a carrier-envelope phase (CEP) zero, spectral peaks scale linearly with THz field strength, characteristic of subcycle emission; (2) At nearly opposite CEP, dominant deceleration fields generate stationary low-energy peaks. Crucially, we develop a pump-probe-free, direct reconstruction method extracting electron pulse profiles solely from measured energy spectra, obtaining durations from 73.0 to 81.0 fs as the field increases (191-290 kV/cm). Phase-resolved simulations further reveal a 72.8% modulation in the cutoff energy and a near-total (99.7%) suppression of the emission current. This work not only validates the field-emisssion theory under THz excitation but also establishes a general framework for the direct temporal characterization of subcycle electron emission, opening pathways for precise electron control in ultrafast electron sources and lightwave nanoelectronics.

Ultrafast electron sources have become indispensable tools to probe matter at fundamental time and length scales, enabling breakthroughs across diverse areas including time-resolved electron microscopy [1], lightwave electronics [2], and coherent radiation sources [3,4]. A critical challenge in these applications lies in achieving precise control over electron emission and acceleration at subcycle timescales [5,6] – a regime that underlies strong-field phenomena [7,8] and delineates the quantum-to-classical transition during electron tunneling [9,10].

Recent advances in laser-driven electron emission have revealed rich dynamics ranging from multiphoton processes to field emission [11-13]. Quantitatively, their division is characterized by the Keldysh parameter [14] $\gamma = \sqrt{\phi/2U_P}$, where $\phi$ is the work function, and $U_P$ denotes the ponderomotive energy that is proportional to the laser intensity and the reciprocal of the angular frequency squared. A value of $\gamma < 1$ (low frequency and strong field regime) corresponds to quasistatic tunneling. In this regime, the instantaneous emission current are well-described by the Fowler-Nordheim (FN) model [15] and its more rigorous extensions that account for the image-charge effect (Murphy-Good (MG) equation) [16]:

$$J = \frac{a(\beta E)^2}{t_F^2 \phi} \exp\left(-\frac{v_F b \phi^{3/2}}{\beta E}\right), \tag{1}$$

Here, $J$ is the emission current, $\beta$ is the field enhancement factor, $E$ is the incident optical field, and $a$, $b$ are FN constants. The correction factors $v_F$ and $t_F$ depend on the scaled barrier field and quantitatively capture the barrier lowering due to the image charge (see Supplementary Section 2 [17] for a detailed discussion).

The temporal characterization of ultrafast emission - a phenomenon essentially equivalent to the first step of the famous three-step model [21] in strong field physics - faces inherent difficulties due to the spectral overlapping of photoelectrons. Particularly, in uniform fields, direct energy-to-time mapping is inhibited by the nonmonotonic photoelectron spectra as well as the trajectory degeneracy where 'short' and 'long' classical paths could yield identical final energies [22,23]. These constraints necessitate pump-probe techniques like attosecond streaking [7,24] and quantum trajectory selection [25,26] to temporally resolve phase-dependent contribution in high harmonic generation or photoelectron spectroscopy, which methods, however, demand exquisite spatiotemporal synchronization between the driving field and probe pulses.

For solid-state electron sources, nanostructured field emitters - particularly sharp nanotips - have emerged as promising platforms due to their ability to concentrate optical fields [27]. The spatial adiabaticity parameter $\delta = l_F/l_q$ further classifies electron dynamics into distinct regimes [28]: (1) when $\delta \geq 1$, electron quiver amplitude $l_q$ fall within the near-field decay length $l_F$, resulting in complex rescattering processes (quiver regime); (2) Otherwise, when $\delta < 1$, the spatial inhomogeneity of near fields naturally filters rescattering electrons, pushing dynamics deeper into the subcycle regime. This spatial-reliant effect creates conditions where electron spectra become dominated by direct electrons - a prerequisite for reliable temporal reconstruction.

Our work exploits the unique advantages of single-cycle, carrier-envelope phase (CEP) stabilized terahertz (THz) pulses for subcycle field emission studies. The long wavelength of the THz pulse ($\lambda \approx 185\ \mu\mathrm{m}$) enables deeper penetration into the subcycle regime through the $\delta \propto \omega^2$ scaling ($\omega$ is the optical frequency), while the single-cycle waveform guarantees isolated emission windows. Crucially, the THz near-field creates a monotonic energy-time mapping where earlier-emitted electrons experience stronger acceleration, which contradicts the conventional Simpleman scenarios [29]. This unique relationship, combined with suppressed rescattering in spatially decaying fields, enables direct reconstruction of emission temporal profiles from electron energy spectra.

While THz-driven subcycle emission [30] has been evidenced via asymmetric emission threshold with flipped field polarity [31] or the linear energy peak shift [32-34] with varying electric field strengths, unambiguous temporal dynamics have remained elusive. Here we address this issue through systematic investigation of graphite tip emission driven by intense single-cycle THz pulses. Our study reveals: First, highly phase-dependent emission characteristics show linear scaling at $\varphi = 0\pi$, whereas an unreported stationary low-energy peak remains at $\varphi = 0.7\pi$. Second, a direct time-domain characterization method reconstructs the temporal electron profiles, which validates the FN theory by demonstrating a pulse width varying from 73.0 to 81.0 fs with increasing field intensity. Third, a phase-resolved simulation demonstrates a remarkable 72.8% CEP modulation of electron cutoff energies and a near-total (99.72%) suppression of the emission current. These advances provide both fundamental insights into subcycle emission physics and practical tools for ultrafast electron control.

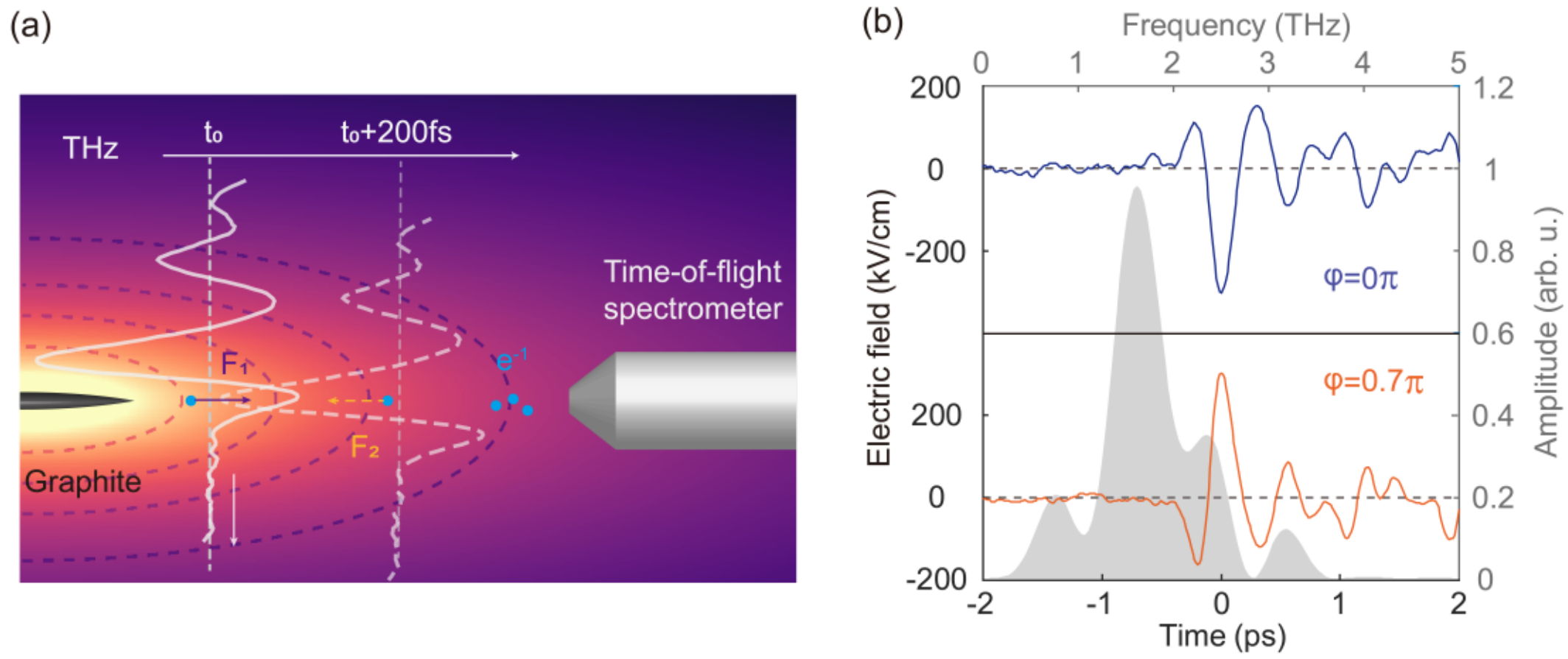

FIG. 1. THz-driven electron dynamics and field characterization. (a) THz-driven subcycle emission from a graphite tip: Negative THz half-cycle triggers electron emission (blue) and accelerates (purple $F_1$) electrons, while the positive half-cycle induces deceleration (yellow $F_2$). For low-energy electrons, the less confined inhomogeneous near-field causes sustained interaction with both THz half-cycles for hundreds of femtoseconds. (b) Measured THz waveforms via electro-optic sampling, with a peak field of 328 kV/cm and the corresponding Fourier spectrum shown in gray shading.

The investigation of subcycle electron emission dynamics demands control over both emission timing and acceleration fields with precise phase stability. As illustrated in Fig. 1(a) and Supplementary Section 1 [17], this control is realized in our experiment through the use of quasi-single-cycle THz pulses generated in a DSTMS crystal [35,36]. The collinear geometry of the rectification process provides inherent CEP stability and facilitates straightforward THz beam alignment. These pulses are then focused in ultrahigh vacuum ($1 \times 10^{-5}$ Pa) onto a graphite tip (electron microscope images in Fig. S1), where they generate enhanced near-fields capable of driving electrons across the tunneling barrier within femtosecond timescales. The emitted electrons are characterized using a source meter (Keithley 6430) accompanied by a time-of-flight (TOF) energy spectrometer (Kaesdorf ETF10), with a static bias of -20 V maintained on the tip throughout measurements.

We first characterized the electron dynamics by their CEP-dependence of the emission yields as a function of THz electric field strength. By inverting the DSTMS crystal upside down, a decisive CEP phase shift between $0\pi$ and $0.7\pi$ is enabled which spans nearly half the THz cycle. The field polarity was determined via THz streaking [37,38] of the electrons (see Supplementary Section 1 [17]). For our tip geometry and THz wavelength, far-to-near-field phase shifts are negligible (see Supplementary Section 5 [17]), ensuring that the CEP measured in the far field accurately represents the phase driving the emission dynamics. Since the THz waveform has a measured peak ratio of 0.53 between negative and positive half-cycles (see Fig. 1(b)), distinct electron emission thresholds were observed similar to the previous observation. As presented in Fig. 2(a), at $\varphi = 0\pi$, a stronger negative field dominates the emission field (see the purple arrow $F_1$ of Fig. 2(a)) which gives rise lower emission threshold of 137 kV/cm, whereas the $\varphi = 0.7\pi$ case exhibits a slightly higher value of 180 kV/cm as a result of weaker extracting electric field force $F_2$ (yellow arrow of Fig. 2(a)).

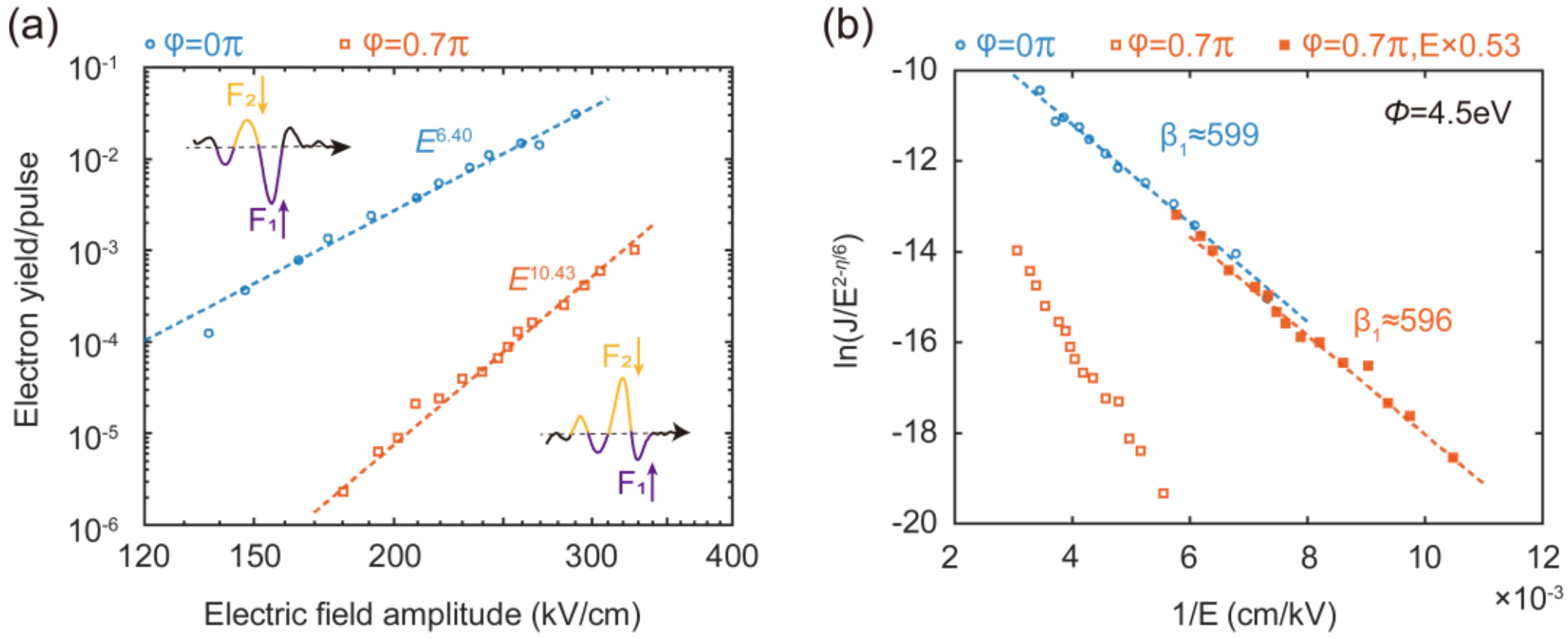

FIG. 2. Phase-dependent electron emission. (a) Measured electron emission yields as a function of the THz field strengths at $\varphi = 0\pi$ (exponent 6.40) and $\varphi = 0.7\pi$ (exponent 10.43). Insets: THz-waveform-driven electron acceleration (purple, $F_1$) and deceleration (yellow, $F_2$). (b) MG plots of the electron yields. The orange solid squares represent the field-scaled data of $\varphi = 0.7\pi$ (orange hollow squares) by the peak ratio of 0.53. The local field enhancement factor is approximately 599 from numerical simulation.

To quantitatively confirm the tunneling emission mechanism, we analyze the current-field characteristic using a transformation of equation (1) [39]:

$$\ln\left(\frac{J}{E^{2-\eta/6}}\right) = \ln\left(A_f^{\mathrm{SN}} \cdot \theta \exp(\eta) \cdot (\beta/E_c)^{2-\eta/6}\right) - \frac{b\phi^{3/2}}{\beta E}, \qquad (2)$$

where $A_f^{\mathrm{SN}}$ is the the formal emission area, $\theta = ac_s^{-4}\phi^3 = aE_c^2\phi^{-1}$ and $\eta = bc_s^2\phi^{-1/2}$ are the scaling parameters, and $c_s = (e^3/4\pi\varepsilon_0)^{1/2}$ is the Schottky constant. The equation (2) stipulates a linear scaling of the quantity $\ln\left(\frac{J_{FN}}{E^{2-\eta/6}}\right)$ with the reciprocal of the applied electric field $E$. We verified this linear relation in Fig. 2(b), which yields a local field enhancement factor of $\beta \approx 599$. The spatial adiabaticity parameter $\delta$, calculated using a near-field decay length of $l_f \approx 312$ nm derived from our tip geometry, falls in the range of 0.54 to 1.14 across our experimental field strengths, confirming the transition into the subcycle, spatially non-adiabatic regime ($\delta < 1$) at higher fields.

Intuitively, a deeper insight into electron dynamics should incorporate the electron's kinetic energy because the time-accumulated action of the optical field could be (implicitly or explicitly) reflected in its final state. Hence, we next investigate the kinetic energies of these emitted electrons. The normalized spectral distributions are presented in Figs. 3 (a) and (b) for $\varphi = 0\pi$ and $\varphi = 0.7\pi$, respectively. While their shapes both consist of a single peak with an extended pedestal, the two CEP cases exhibit stark contrasts in terms of peak position and the pedestal's extension directions. As demonstrated in Fig. 3 (c), the spectral peaks at $\varphi = 0\pi$ displayed linear scaling with THz field strength - a characteristic signature of subcycle emission. Conversely, at $\varphi = 0.7\pi$, a stationary low-energy peak was observed near 21.5 eV regardless of applied field intensity. The stationary peak is accompanied by a plateau-like high-energy pedestal, which, however, differs fundamentally from uniform-field photoelectron spectra.

To elucidate the physical origin of these spectral features, we numerically simulate electron tunneling and propagation based on a semiclassical two-step model (see Supplementary Section 2 [17]). The simulation incorporates the realistic emitter geometry, which exhibits nanoscale protrusions (Fig. S1) governing the local field enhancement, while the overall tip geometry affects subsequent acceleration. First, electron tunneling through the field-lowered barrier is treated quantum-mechanically, with the instantaneous emission probability described by the MG equation. Subsequently, the emitted electrons are propagated as classical particles in the combined THz and static fields. Notably, this detailed two-field description can be represented by a simplified effective model [40-42], which provides a more straightforward picture while yielding consistent key results (see Supplementary Section 4 [17]), underscoring the generality of the observed dynamics. The electron trajectories were computed for ensembles of electrons with initial emission times weighted by the instantaneous tunneling rate. Further details of the simulation are provided in the Supplementary Section 2 [17].

The simulations yield key spectral features of the experiment. In Figs. 3 (a) and (b), it is evident that the characteristic peak shifts in the experiment (blue solid lines) are well reproduced in the simulation (red dashed lines), while the calculated peak for $\varphi = 0.7\pi$ shows much narrower widths approaching ~0.2 eV. To illustrate the result, we first provide a explanation of the unconventional anchored peak of the electron energy. While unfolding the final kinetic energy of an emitted electron as a function of its emission time (see Fig. S3), the spectrum is essentially a convolution of field-dependent tunneling probability with the acceleration dynamics. In the absence of (or with a weak) ensuing deceleration THz field such as at $\varphi = 0\pi$, the electron gains energy approximately proportional to the instantaneous field strength, thereby creating the stiff precipice at the high energy edge due to a maximal tunneling rate. However, when subsequent deceleration is strong, backward accelerations redistribute most electrons toward the detection threshold (which in our case is set mainly by the DC acceleration voltage), consequently forming anchored peaks. In this scenario, the observed spectral broadening, particularly at $\varphi = 0.7\pi$, stems primarily from the spatial inhomogeneity of the THz near-field, which suppresses high-energy rescattering plateaus, and the energy-dependent transmission function of the TOF spectrometer. These experimental factors, detailed in Supplementary Section 3 [17], account for the differences with the idealized simulation.

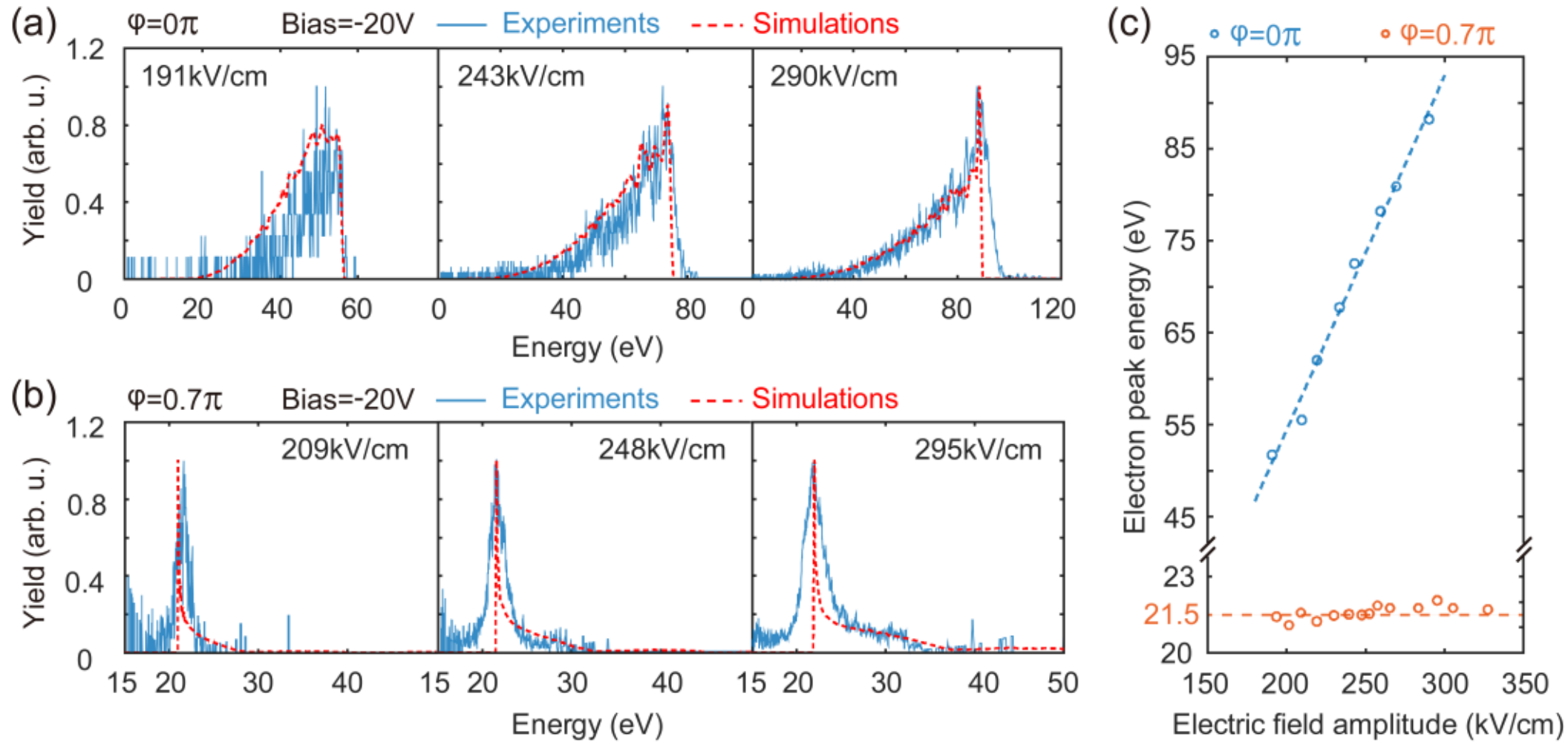

FIG. 3. Phase-dependent electron energy spectra. (a) Measured (blue) and simulated (red) spectra at $\varphi = 0\pi$ for THz field strengths of 191 kV/cm, 243 kV/cm, and 290 kV/cm. (b) Corresponding data at $\varphi = 0.7\pi$ for similar THz field strengths of 209 kV/cm, 248 kV/cm, and 295 kV/cm. The simulated electron spectra in (a) and (b) are presented with resolution

matching the TOF's measurement precision. (c) Field-dependent peak energy scaling.

A further inspection of the time-energy mapping provides critical insights into the characteristic timescales governing subcycle electron dynamics. Figures 4(a) and (b) present a detailed view of simulated emission time versus final kinetic energy for $\varphi = 0\pi$ and $\varphi = 0.7\pi$, respectively. To extend the analysis beyond the experimentally accessible THz field amplitudes, we modulate the spatial adiabaticity parameter $\delta$ in simulations by varying the THz field strengths from 30 to 1000 kV/cm, with the experimental parameter range highlighted by the shaded area. Notably, two field-dependent effects emerge as $\delta$ decreases with increasing THz field strength: First, the emission time (phase) $\phi_0$ corresponding to the maximum energy of directly emitted electrons progressively shifts towards the peak of the THz field (located at $t \approx 0$ for $\varphi = 0\pi$ and $t \approx -197$ fs for $\varphi = 0.7\pi$, where $t = 0$ denotes the THz envelope maximum). Second, the emission time of minimum energy is postponed, and the successive deceleration field could induce rescattering (colored dashed lines in Figs. 4(a) and (b)).

The peak shifts have direct implications for the spectral distributions. Specifically, at $\varphi = 0\pi$ (Fig. 4(a)), the emission time ($\phi_0$) for maximum electron energy moves closer to $t = 0$ as the field intensity increases. Simultaneously, the slope of the final kinetic energy versus emission time steepens near $t = 0$. These combined effects cause more electrons to concentrate at higher energies, resulting in the sharper high-energy spectral cutoffs observed experimentally in Fig. 3(a). In contrast, at $\varphi = 0.7\pi$ (Fig. 4(b)), rescattering processes dominate the highest energy portion of the spectrum. Notably, the peak positions of the direct electron energies stay distant from the THz field peak location, and the low-energy peak near $t \approx -197$ fs remains remarkably stable across our experimental THz field range. This stability directly accounts for the stationary spectral peak observed in Fig. 3(b).

The field-induced shifts in the energy-time mapping create a dynamic window exhibiting a monotonic dependence of final kinetic energy on emission time, which allows for direct reconstruction of temporal emission profiles from energy spectra. Crucially, the quasi-single-cycle nature of the THz waveform ensures a well-defined, isolated emission window. The case at $\varphi = 0\pi$ (shown conceptually in Fig. S3(a)) exemplifies this ideal condition for temporal reconstruction: as Fig. 4(c) illustrates, the monotonic energy-time correspondence enables a direct mapping over the range [-24.6, 66.7] fs (colored shading), which captures the majority of emitted electrons.

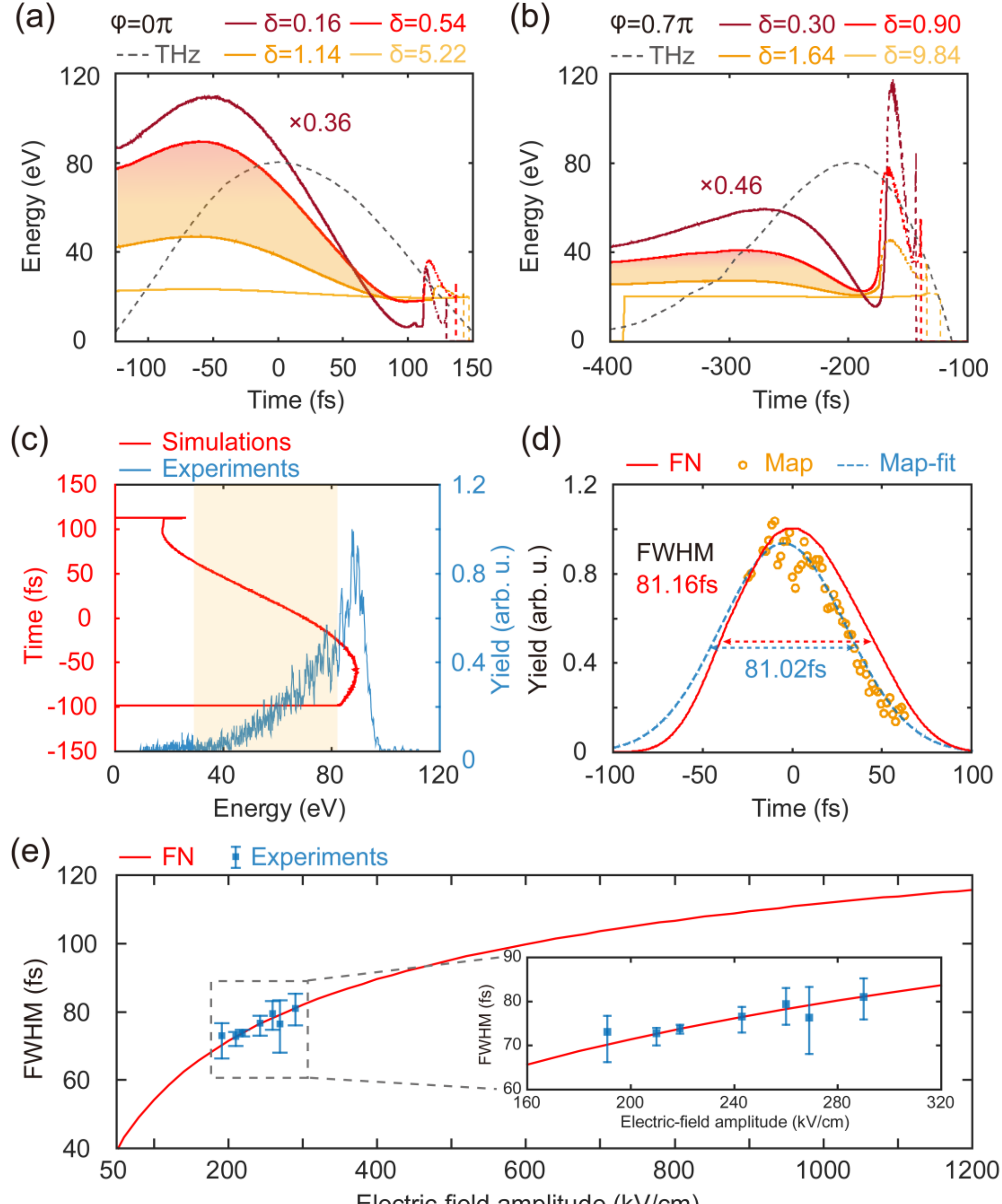

FIG. 4. Time-Energy Correlation of THz-Driven Electron Emission. (a) and (b), simulated final electron energy versus emission time. Solid and dashed colored lines denote direct and rescattering electrons, respectively, under different THz field strengths. The gray dashed line represents the THz electric field. The light red shaded region marks the experimental parameter range ($t = 0$: THz envelope maximum). (c) Simulated emission time–energy correlation (red) and detected electron energy spectrum (blue). (d) A mapping result of the electron emission profile with 290 kV/cm THz. (e) Mapped electron pulse widths (blue) are consistent with MG

equation predictions (red). The error bars include the energy jitter of our THz source, the fitting errors of the energy spectrum and the mapping process.

Leveraging this mapping, we reconstructed the temporal electron emission profile directly from the experimental energy spectrum. The yellow data points in Fig. 4(d) represent this reconstructed profile. The full width at half maximum (FWHM) of the emission pulse was determined to be 81.0 (-5.1, +4.2) fs, demonstrating close agreement with the 81.2 fs prediction derived from the MG equation. The robustness of this direct reconstruction method against the detailed emitter morphology is confirmed through comprehensive simulations, including a Monte Carlo analysis of distributed nanoscale emitters (see Supplementary Section 4 [17]).

Figure 4(e) reveals the dependence of the reconstructed pulse width on the applied THz field. As the THz field strength increases from 191 kV/cm to 290 kV/cm, the pulse width systematically broadens from 73.0 (-6.9, +3.7) fs to 81.0 (-5.1, +4.2) fs. This pulse broadening is a hallmark of the tunneling regime, distinct from the pulse shortening typical of multiphoton emission. In the tunneling picture, a higher peak field exponentially increases the emission probability but also extends the temporal window where the tunneling rate is significant. The error bars in Fig. 4(e) incorporate uncertainties arising from spectral fitting, laser energy fluctuations, and detection precision, underscoring the robustness of our reconstruction method.

Essentially, the energy-time correlations in Figs. 4(a) and (b), and their resultant spectral features, stem from the combined spatiotemporal asymmetry inherent to our experiment. This is distinct from effects dominated solely by spatial field gradients in multi-cycle pulses [43,44]. In our regime, the vector potential of the single-cycle field does not return to zero at the field zero-crossing [45], fundamentally breaking the time-reversal symmetry of the electron acceleration process. When this temporal asymmetry is combined with the spatial gradient of the near-field, it suppresses rescattering, shifts saddle points of energy-time correlation, and creates complex phase-space filtering of electrons that produce the sloping-shaped and the peak-anchored energy spectra in Figs. 3(a) and (b), a feature qualitatively distinct from those with multicycle pulses [43,46].

This strong phase dependence of electron emission is expected to significantly influence the final energy spectra. To comprehensively characterize this THz-field-dependent energy modulation, we simulated the energy spectra across varying CEP values (see THz waveforms in Fig. S7). Numerical simulations of CEP-modulated electron spectra in Fig. 5(a) confirm the strong CEP dependence of electron dynamics, with particularly pronounced effects contrasting $\varphi_{\mathrm{CEP}} = 0.66\pi$ and $\varphi_{\mathrm{CEP}} = 1.64\pi$, which correspond to the strongest net acceleration and deceleration fields, respectively, for our non-Gaussian asymmetric THz envelope (see Fig. S7). The cutoff energy in Fig. 5(b), defined statistically as the energy below which 95% of the total electron population resides, exhibited a theoretical modulation depth of 72.8% at these two phases. Experimentally, this depth was measured as 57.1% at our applied CEPs, aligning well with the simulated value of 60.5%. The large contrast originates from the electron velocity evolution (Fig. 5(c)), where distinct deceleration for electrons emitted at $\varphi_{\mathrm{CEP}} = 0.66\pi$ results in a minimum energy peak. The CEP results reveal a substantial 99.72% reduction in emission number. Since this emission energy and charge density modulation is singly reliant on the CEP, the observed high modulation depth demonstrates significant potential for precise, femtosecond-scale control of ultrafast electron pulses.

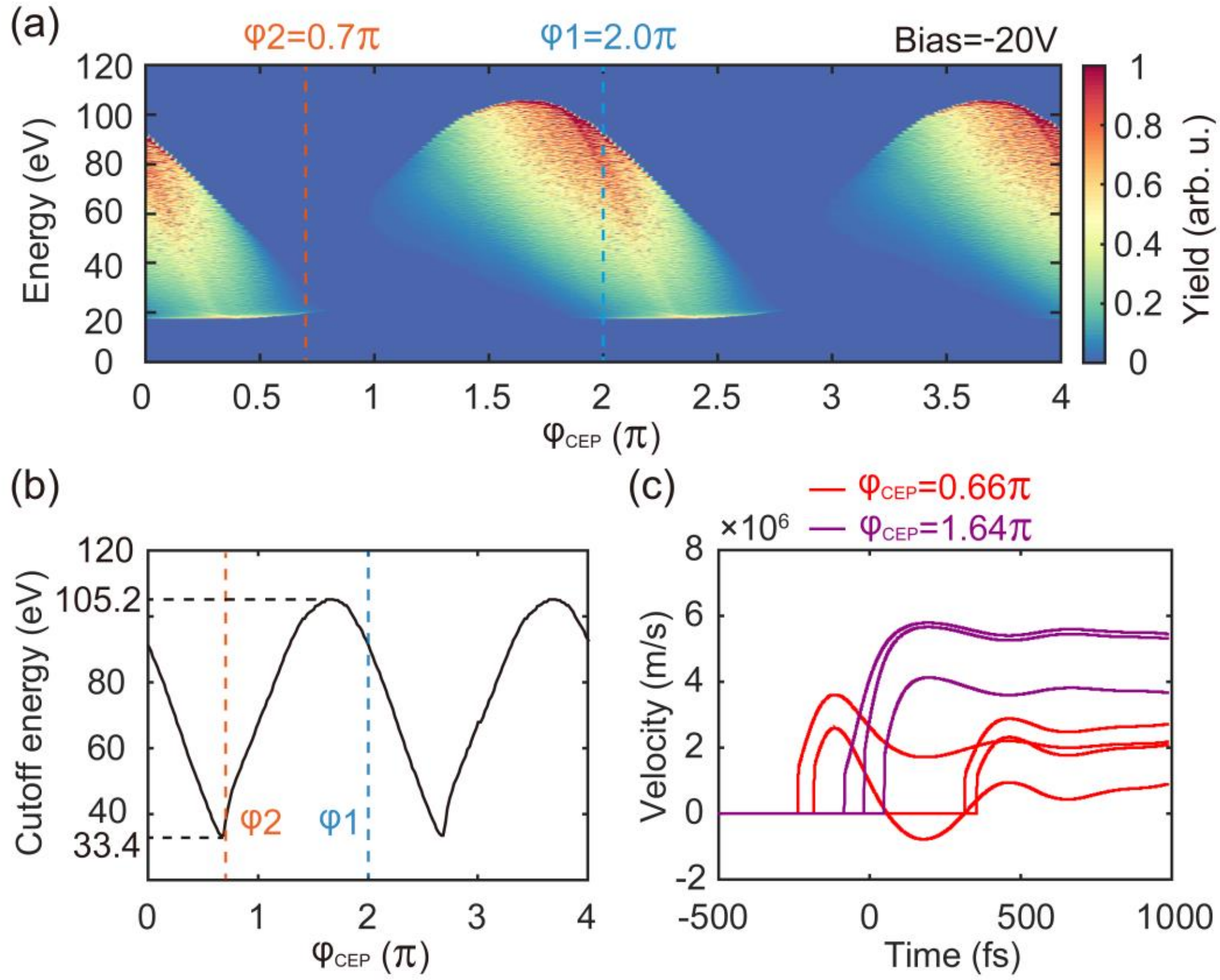

FIG. 5. CEP modulation of electron emission dynamics. (a) Simulated electron energy spectra as a function of THz CEP $\varphi_{\mathrm{CEP}}$. Spectra corresponding to the experimental phases $\varphi_1$ (blue) and $\varphi_2$ (red) are indicated. (b) CEP-dependent high-energy cutoff, statistically defined as the energy below which 95% of the total electron population resides. The theoretical modulation depth reaches 72.8%. (c) Simulated electron velocity evolution profiles versus time delay

within the spatially inhomogeneous near field, demonstrating distinct acceleration/deceleration behaviors dependent on the emission phase.

To sum up, our investigation of carrier-envelope-phase-stabilized, single-cycle terahertz-driven electron emission from a graphite tip reveals distinct phase-dependent dynamics critical for subcycle temporal characterization. At a zero phase of $\varphi = 0\pi$, the linear scaling of spectral peak energy with THz field strength is observed, characteristic of subcycle emission governed by the FN model. Crucially, the THz near-field configuration at this phase establishes a monotonic mapping between electron kinetic energy and emission time, a direct consequence of the accelerating field profile where earlier-emitted electrons gain higher final energies. This unique relationship, arising from the single-cycle waveform and suppressed rescattering in the inhomogeneous near-field, enables direct reconstruction of the temporal emission profile solely from the measured energy spectrum - a concept also demonstrated in spatially inhomogeneous mid-infrared fields [47]. It proves robustness across different theoretical descriptions of field emission and is insensitive to the detailed nanoscale emitter morphology (see Supplementary Section 4 [17]), as confirmed by the two-field model analysis. Applying this method, we determined unambiguous subcycle emission pulse widths ranging from 73.0 to 81.0 fs as the THz field increased from 191 kV/cm to 290 kV/cm, corroborating the predicted trend from field emission theory.

The subcycle emission is strongly influenced by the CEP. Aside from the stationary peak near 21.5 eV at $\varphi = 0.7\pi$, our phase-resolved simulations revealed a 72.8% modulation depth in electron cutoff energy. At higher driving THz fields, the emission is expected to penetrate deeper into the subcycle regime, enabling greater modulation depth and cutoff energy exceeding the keV range. Given the temporal demonstrates an optically driven, ultrafast electron source capable of delivering ~100 fs pulse widths with the majority of electrons concentrated in the high-energy region. We note that THz-driven electron emission exceeding the keV range has been previously reported [32-34,48]; the timescales of such electrons could be directly retrieved using our reconstruction approach.

The reconstruction approach could also be extended to other wavelengths beyond THz with a single emission window. Though the temporal uncertainty of this approach is approximately 7 fs in our experiment, we underscore the precision of this method is essentially constrained by three factors: (1) the energy resolution of the time-of-flight spectrometer; (2) The monotonic correlation window width; and (3) the data fineness of the measured energy spectrum. The error of our current experiment arises mainly from the third factor, which suffers from a coarse spectral curve due to a low collection efficiency of the TOF. However, the developed direct time-domain characterization method, leveraging energy-time mapping in tailored THz near-fields, establishes a powerful approach for femtosecond-scale temporal profiling and manipulation of electron pulses, with significant implications for advancing ultrafast electron sources and lightwave-driven nanoelectronics.

Acknowledgments-We are grateful to Zhinan Zeng from the Zhangjiang Laboratory, Songyu Zhu and Ruoyu Chen from SIOM for previous discussions. This work was supported by the National Key Research and Development Program of China (2022YFA1604400); National Natural Science Foundation of China (12325409, 12388102, U2267204); Shanghai Pilot Program for Basic Research, Chinese Academy of Sciences, Shanghai Branch, CAS Project for Young Scientists in Basic Research (YSBR-060).

Note - After the submission of this work, we became aware of a highly related study exploring near-field-induced low-energy spectral features in the attosecond domain [Heimerl et al., arXiv:2507.02673].

*These authors contributed equally to this work.
†Contact Author: slw@siom.ac.cn
‡Contact Author: tianye@siom.ac.cn
§Contact Author: ruxinli@siom.ac.cn

## References

[1] D. Nabben, J. Kuttruff, L. Stolz, A. Ryabov, and P. Baum, Attosecond electron microscopy of sub-cycle optical dynamics, Nature **619**, 63 (2023).

[2] M. Borsch, M. Meierhofer, R. Huber, and M. Kira, Lightwave electronics in condensed matter, Nat. Rev. Mater. **8**, 668 (2023).

[3] D. Zhang, Y. Zeng, Y. Tian, and R. Li, Coherent free-electron light sources,

Photon. Insights **2**, R07 (2023).

[4] D. Zhang, Y. Zeng, Y. Bai, Z. Li, Y. Tian, and R. Li, Coherent surface plasmon polariton amplification via free-electron pumping, Nature **611**, 55 (2022).

[5] X. Yu, Y. Zeng, L. Song, D. Kong, S. Hao, J. Gui, X. Yang, Y. Xu, X. Wu, Y. Leng, Y. Tian, and R. Li, Megaelectronvolt electron acceleration driven by terahertz surface waves, Nat. Photonics **17**, 957 (2023).

[6] D. Zhang, A. Fallahi, M. Hemmer, X. Wu, M. Fakhari, Y. Hua, H. Cankaya, A.-L. Calendron, L. E. Zapata, N. H. Matlis, and F. X. Kärtner, Segmented terahertz electron accelerator and manipulator (STEAM), Nat. Photonics **12**, 336 (2018).

[7] H. Y. Kim, M. Garg, S. Mandal, L. Seiffert, T. Fennel, and E. Goulielmakis, Attosecond field emission, Nature **613**, 662 (2023).

[8] P. Dienstbier, L. Seiffert, T. Paschen, A. Liehl, A. Leitenstorfer, T. Fennel, and P. Hommelhoff, Tracing attosecond electron emission from a nanometric metal tip, Nature **616**, 702 (2023).

[9] P. Dombi, Z. Pápa, J. Vogelsang, S. V. Yalunin, M. Sivis, G. Herink, S. Schäfer, P. Groß, C. Ropers, and C. Lienau, Strong-field nano-optics, Rev. Mod. Phys. **92**, 025003 (2020).

[10] R. Bormann, M. Gulde, A. Weismann, S. V. Yalunin, and C. Ropers, Tip-Enhanced Strong-Field Photoemission, Phys. Rev. Lett. **105**, 147601 (2010).

[11] B. Bánhegyi, G. Z. Kiss, Z. Pápa, P. Sándor, L. Tóth, L. Péter, P. Rácz, and P. Dombi, Nanoplasmonic Photoelectron Rescattering in the Multiphoton-Induced Emission Regime, Phys. Rev. Lett. **133**, 033801 (2024).

[12] G. Hergert, R. Lampe, and C. Lienau, Quenching Strong-Field Rescattering of Photoemitted Electrons from Metallic Nanotapers by Using Moderate Bias Fields, ACS Photonics **12**, 2219 (2025).

[13] J. Schötz, S. Mitra, H. Fuest, M. Neuhaus, W. A. Okell, M. Förster, T. Paschen, M. F. Ciappina, H. Yanagisawa, P. Wnuk, P. Hommelhoff, and M. F. Kling, Nonadiabatic ponderomotive effects in photoemission from nanotips in intense midinfrared laser fields, Phys. Rev. A **97**, 013413 (2018).

[14] L. V. Keldysh, Ionization in the field of a strong electromagnetic wave, J. Exp.

Theor. Phys. **20**, 1307 (1965).

[15] R. H. Fowler, and L. Nordheim, Electron emission in intense electric fields, Proc. R. Soc. Lond. A **119**, 173 (1928).

[16] E. L. Murphy, and R. H. Good, Thermionic Emission, Field Emission, and the Transition Region, Physical Review **102**, 1464 (1956).

[17] See Supplemental Material [url] for experimental setup, two-field numerical simulation model, origin of spectral broadening in the experiment, robustness of temporal reconstruction, and far-to-near-field CEP shift at THz frequencies, which includes Refs. [16, 18-20, 39-42].

[18] R. G. Forbes, Simple good approximations for the special elliptic functions in standard Fowler-Nordheim tunneling theory for a Schottky-Nordheim barrier, Appl. Phys. Lett. **89**, 113122 (2006).

[19] L. Wimmer, O. Karnbach, G. Herink, and C. Ropers, Phase space manipulation of free-electron pulses from metal nanotips using combined terahertz near fields and external biasing, Phys. Rev. B **95**, 165416 (2017).

[20] K. Yoshioka, I. Katayama, Y. Arashida, A. Ban, Y. Kawada, K. Konishi, H. Takahashi, and J. Takeda, Tailoring Single-Cycle Near Field in a Tunnel Junction with Carrier-Envelope Phase-Controlled Terahertz Electric Fields, Nano Lett. **18**, 5198 (2018).

[21] P. B. Corkum, Plasma perspective on strong field multiphoton ionization, Phys. Rev. Lett. **71**, 1994 (1993).

[22] D. Shafir, H. Soifer, B. D. Bruner, M. Dagan, Y. Mairesse, S. Patchkovskii, M. Y. Ivanov, O. Smirnova, and N. Dudovich, Resolving the time when an electron exits a tunnelling barrier, Nature **485**, 343 (2012).

[23] N. Dudovich, O. Smirnova, J. Levesque, Y. Mairesse, M. Y. Ivanov, D. M. Villeneuve, and P. B. Corkum, Measuring and controlling the birth of attosecond XUV pulses, Nat. Phys. **2**, 781 (2006).

[24] S. Brennecke, M. Ranke, A. Dimitriou, S. Walther, M. J. Prandolini, M. Lein, and U. Frühling, Control of Electron Wave Packets Close to the Continuum Threshold Using Near-Single-Cycle THz Waveforms, Phys. Rev. Lett. **129**, 213202 (2022).

[25] A. J. Piper, Q. Liu, A. Camacho Garibay, D. Kiesewetter, V. Leshchenko, J. E. Bækhøj, P. Agostini, K. J. Schafer, L. F. DiMauro, and Y. Tang, Attosecond Clocking and Control of Strong Field Quantum Trajectories, Phys. Rev. Lett. **134**, 073201 (2025).

[26] K. J. Schafer, M. B. Gaarde, A. Heinrich, J. Biegert, and U. Keller, Strong Field Quantum Path Control Using Attosecond Pulse Trains, Phys. Rev. Lett. **92**, 023003 (2004).

[27] D. Ehberger, J. Hammer, M. Eisele, M. Krüger, J. Noe, A. Högele, and P. Hommelhoff, Highly Coherent Electron Beam from a Laser-Triggered Tungsten Needle Tip, Phys. Rev. Lett. **114**, 227601 (2015).

[28] G. Herink, D. R. Solli, M. Gulde, and C. Ropers, Field-driven photoemission from nanostructures quenches the quiver motion, Nature **483**, 190 (2012).

[29] F. Krausz, and M. Ivanov, Attosecond physics, Rev. Mod. Phys. **81**, 163 (2009).

[30] G. Herink, L. Wimmer, and C. Ropers, Field emission at terahertz frequencies: AC-tunneling and ultrafast carrier dynamics, New J. Phys. **16**, 123005 (2014).

[31] S. Li, A. Sharma, Z. Márton, P. S. Nugraha, C. Lombosi, Z. Ollmann, I. Márton, P. Dombi, J. Hebling, and J. A. Fülöp, Subcycle surface electron emission driven by strong-field terahertz waveforms, Nat. Commun. **14**, 6596 (2023).

[32] S. Li, and R. R. Jones, High-energy electron emission from metallic nano-tips driven by intense single-cycle terahertz pulses, Nat. Commun. **7**, 13405 (2016).

[33] D. Matte, N. Chamanara, L. Gingras, L. P. R. de Cotret, T. L. Britt, B. J. Siwick, and D. G. Cooke, Extreme lightwave electron field emission from a nanotip, Phys. Rev. Res. **3**, 013137 (2021).

[34] B. Colmey, R. T. Paulino, G. Beaufort, and D. G. Cooke, Sub-cycle nanotip field emission of electrons driven by air plasma generated THz pulses, Appl. Phys. Lett. **126**, 031108 (2025).

[35] K. Wang, Z. Zheng, H. Li, X. Meng, Y. Liu, Y. Tian, and L. Song, Efficient strong-field THz generation from DSTMS crystal pumped by 1030 nm Yb-laser, Appl. Phys. Lett. **124**, 121102 (2024).

[36] Z. Zheng, K. Wang, H. Li, X. Meng, Y. Tian, and L. Song, High-repetition-rate

strong-field terahertz source by optical rectification in DSTMS crystals, High Power Laser Sci. Eng. **12**, 05000e61 (2024).

[37] L. Wimmer, G. Herink, D. R. Solli, S. V. Yalunin, K. E. Echternkamp, and C. Ropers, Terahertz control of nanotip photoemission, Nat. Phys. **10**, 432 (2014).

[38] C. Kealhofer, W. Schneider, D. Ehberger, A. Ryabov, F. Krausz, and P. Baum, All-optical control and metrology of electron pulses, Science **352**, 429 (2016).

[39] R. G. Forbes, The Murphy–Good plot: a better method of analysing field emission data, R. Soc. Open Sci. **6**, 190912 (2019).

[40] T. L. Cocker, V. Jelic, M. Gupta, S. J. Molesky, J. A. J. Burgess, G. D. L. Reyes, L. V. Titova, Y. Y. Tsui, M. R. Freeman, and F. A. Hegmann, An ultrafast terahertz scanning tunnelling microscope, Nat. Photonics **7**, 620 (2013).

[41] K. Iwaszczuk, M. Zalkovskij, A. C. Strikwerda, and P. U. Jepsen, Nitrogen plasma formation through terahertz-induced ultrafast electron field emission, Optica **2**, 116 (2015).

[42] J. C. Blancon, A. V. Stier, H. Tsai, W. Nie, C. C. Stoumpos, B. Traoré, L. Pedesseau, M. Kepenekian, F. Katsutani, G. T. Noe, J. Kono, S. Tretiak, S. A. Crooker, C. Katan, M. G. Kanatzidis, J. J. Crochet, J. Even, and A. D. Mohite, Scaling law for excitons in 2D perovskite quantum wells, Nat. Commun. **9**, 2254 (2018).

[43] D. J. Park, B. Piglosiewicz, S. Schmidt, H. Kollmann, M. Mascheck, and C. Lienau, Strong Field Acceleration and Steering of Ultrafast Electron Pulses from a Sharp Metallic Nanotip, Phys. Rev. Lett. **109**, 244803 (2012).

[44] D. J. Park, B. Piglosiewicz, S. Schmidt, H. Kollmann, M. Mascheck, P. Groß, and C. Lienau, Characterizing the optical near-field in the vicinity of a sharp metallic nanoprobe by angle-resolved electron kinetic energy spectroscopy, Ann. Phys. **525**, 135 (2013).

[45] S. Li, and R. R. Jones, Ionization of Excited Atoms by Intense Single-Cycle THz Pulses, Phys. Rev. Lett. **112**, 143006 (2014).

[46] B. Piglosiewicz, S. Schmidt, D. J. Park, J. Vogelsang, P. Groß, C. Manzoni, P. Farinello, G. Cerullo, and C. Lienau, Carrier-envelope phase effects on the strong-

field photoemission of electrons from metallic nanostructures, Nat. Photonics **8**, 37 (2014).

[47] K. E. Echternkamp, G. Herink, S. V. Yalunin, K. Rademann, S. Schäfer, and C. Ropers, Strong-field photoemission in nanotip near-fields: from quiver to sub-cycle electron dynamics, Appl. Phys. B **122**, 80 (2016).

[48] J. Ying, X. He, D. Su, L. Zheng, T. Kroh, T. Rohwer, M. Fakhari, G. H. Kassier, J. Ma, P. Yuan, N. H. Matlis, F. X. Kärtner, and D. Zhang, High gradient terahertz-driven ultrafast photogun, Nat. Photonics **18**, 758 (2024).